\documentclass[a4paper]{IEEETran}
\usepackage[T1]{fontenc}
\usepackage[latin9]{inputenc}
\usepackage{color}
\usepackage{amsmath}
\usepackage{amssymb}
\usepackage{stackrel}
\usepackage{graphicx}
\usepackage{esint}
\usepackage[unicode=true,
 bookmarks=false,
 breaklinks=false,pdfborder={0 0 1},backref=false,colorlinks=false]
 {hyperref}
\usepackage{breakurl}

\makeatletter

\usepackage{xcolor}
\usepackage{pdfcolmk}
\usepackage{units}

\usepackage{epsfig}
\usepackage{float}
\usepackage{subfigure}
\usepackage{subfloat}
\usepackage{cite}

\usepackage{miniplot}

\allowdisplaybreaks

\PassOptionsToPackage{normalem}{ulem}
\usepackage{ulem}

  \pdfpageheight\paperheight
\pdfpagewidth\paperwidth

\renewcommand{\thesubsubsection}{\arabic{subsubsection}}

\makeatother

\begin{document}

\title{On the Security of Warning Message Dissemination in Vehicular Ad
Hoc Networks}

\author{Jieqiong Chen, Guoqiang Mao \\
School of Computing and Communications, University of Technology Sydney, NSW 2007, Australia\\
Email: jieqiong.chen@student.uts.edu.au, g.mao@ieee.org
}
\maketitle
\begin{abstract}
Information security is an important issue in vehicular networks as
the accuracy and integrity of information is a prerequisite to satisfactory
performance of almost all vehicular network applications. In this
paper, we study the information security of a vehicular ad hoc network
whose message may be tampered by malicious vehicles. An analytical
framework is developed to analyze the process of message dissemination
in a vehicular network with malicious vehicles randomly distributed
in the network. The probability that a destination vehicle at a fixed
distance away can receive the message correctly from the source vehicle
is obtained. Simulations are conducted to validate the accuracy of
the theoretical analysis. Our results demonstrate the impact of network
topology and the distribution of malicious vehicles on the correct
delivery of a message in vehicular ad hoc networks, and may provide
insight on the design of security mechanisms to improve the security
of message dissemination in vehicular networks.
\end{abstract}


\section{Introduction}

Interest is surging on vehicular networks and Internet-of-vehicles
technologies due to their increasingly important role in improving
road traffic efficiency, enhancing road safety and providing real-time
information to drivers and passengers \cite{Zheng15}. By deploying
wireless communication infrastructure along the roadside (e.g., road-side
units (RSU)), equipping vehicles with on-board communication facilities
(e.g., on-board units (OBU)), and with the assistance of dedicated
short-range communication (DSRC) \cite{Kenny11} and LTE technology,
two wireless communication modes: vehicle-to-infrastructure and vehicle-to-vehicle
communications, are supported in vehicular networks. Through wireless
communications, messages can be disseminated for vehicular network
applications, including safety applications requiring real-time information
about traffic accidents, traffic congestion or obstacles in the road,
and non-safety applications such as offering value-added services
(e.g., digital maps with real-time traffic status) and in-car entertainment
services \cite{Ilarri15}.

Coming together with the convenience and advantage of wireless communications
is the potential security threat that vehicular networks may present
to transportation systems. Different from traditional security settings,
in vehicular networks, information collection and dissemination are
conducted by distributed vehicles. Quite often, information may be
generated by or received from a vehicle that has never been encountered
before. This may render traditional security mechanisms, largely based
on cryptography and key management, or trust management, futile in
vehicular networks. The situation is further exacerbated by the highly
dynamic topology of vehicular networks where the connections may emerge
opportunistically between vehicles and the associated network topology
is constantly changing \cite{Mao09}. All these features of vehicular
networks pose unique challenges for vehicular network security and
make vehicular networks prone to attacks by malicious and/or selfish
attackers who may spread false messages, tamper or drop the received
messages. These security threats are likely to result in severe consequences
like traffic congestion, traffic crash, even loss of lives and must
be thoroughly investigated before vehicular networks can be deployed. 

In this paper, we study information security of vehicular ad hoc networks
(VANETs), where the message may be tampered by malicious vehicles
randomly distributed in the network, by investigating the probability
that a destination vehicle at a fixed distance away can receive the
message correctly from the source vehicle. Specifically, consider
that a vehicle (i.e., the source vehicle) detecting an abnormal situation,
e.g., traffic accident, slippery road and congestion, sends a message
informing other vehicles of the situation. The message is forwarded
from the source vehicle in a multi-hop manner to other vehicles. We
analyze the probability that a vehicle at a fixed distance away, termed
the destination vehicle, can receive the message correctly from the
source vehicle in the presence of malicious vehicles in between, which
may modify the transmitted message. The novelty and major contributions
of this paper are summarized as follows: 
\begin{enumerate}
\item We develop for the first time an analytical framework to model the
process of message dissemination in vehicular ad hoc networks in the
presence of malicious vehicles randomly distributed in the network.
The probability that a message is delivered correctly from the source
vehicle to a destination vehicle at a fixed distance away is analyzed. 
\item Simulations are conducted to establish the accuracy of the analysis.
Using the analysis, relationship is revealed between key parameters
such as the probability of correct message reception and its major
performance-impacting parameters. Discussions are presented on the
impact of network topology and the distribution of malicious vehicles
on secure message delivery in vehicular networks. 
\item Our results may provide insight on the design of security mechanisms,
particularly secure routing algorithms and topology control algorithms,
to improve informations security in vehicular networks.
\end{enumerate}
The rest of this paper is organized as follows: Section \ref{sec:Related-Work}
reviews related work. Section \ref{sec:System-Model-and} introduces
the system model and the problem formation. Theoretical analysis is
presented in Section \ref{sec:Theoretical-Analysis }. In Section
\ref{sec:Simulation-and-Discussion}, we conduct simulations to validate
the accuracy of our analysis and discuss its insight. Section \ref{sec:Conclusion-and-Future}
concludes this paper.

\section{Related Work\label{sec:Related-Work}}

For secure message dissemination in vehicular networks, two major
factors need to be considered: the trustworthiness of each vehicle
and the integrity of the transmitted message. Accordingly, three misbehavior
detection schemes are commonly adopted for secure message dissemination:
entity-centric misbehavior detection scheme, data-centric misbehavior
detection scheme, and a combined use of both. In the following, we
will review the works on these three schemes separately. 

Entity-centric misbehavior detection schemes focus on assessing the
trustworthiness level of each vehicle to filter out the malicious
vehicles. The assessment process is commonly conducted at each vehicle
by monitoring their instantaneous neighbors' behavior. In \cite{Gazdar12},
Gazdar et al. proposed a dynamic and distributed trust model to formalize
a trust relationship between vehicles and filter out malicious and
selfish vehicles. Their trust model is based on the use of a Markov
chain to evaluate the evolution of the trust value. In \cite{Khan15},
instead of allowing all vehicles to assess trustworthiness, Khan et
al. proposed a novel malicious node detection algorithm for VANETs,
which optimizes the selection of assessors to improve the overall
network performance. In \cite{Haddadou15}, Haddadou et al. proposed
a distributed trust model for VANETs, which was motivated by the job
market signaling model. Their trust model is able to gradually detect
all malicious nodes as well as boosting the cooperation of selfish
nodes. In \cite{Sedjelmaci15}, to overcome the challenges of intermittent
and ad hoc monitoring and assessment processes caused by the high
mobility and rapid topology change in vehicular networks, Sedjelmaci
et al. proposed a lightweight intrusion detection framework with the
help of a clustering algorithm, where nodes are grouped into highly
stable clusters so that the monitoring and assessment processes can
be better conducted in a relatively stable environment. 

Data-centric misbehavior detection schemes focus on the consistency
check of the disseminated data to filter out the false data. In \cite{Dietzel13},
Dietzel et al. argued that redundant data forwarding paths are the
most promising technique for effective data consistency check in a
multi-hop information dissemination environment, and proposed three
graph-theoretic metrics to measure the redundancy of dissemination
protocols. In \cite{Raya08}, Raya et al. proposed a framework for
vehicular networks to establish data-centric trust, and evaluated
the effectiveness of four data fusion rules: majority voting, weighted
voting, Bayesian inference and belief propagation based techniques.
In \cite{Huang14}, Huang et al. firstly demonstrated that information
cascading and oversampling adversely affect the performance of trust
management scheme in VANETs, and then proposed a novel voting scheme
that taking the distance between the transmitter and receiver into
account when assigning weight to the trust level of the received data.
In \cite{Zaidi16}, Zaidi et al. proposed and evaluated a rogue node
detection system for VANETs using statistical techniques to determine
whether the received data are false. In \cite{Radak}, Radak applied
a so-called cautious operator to deal with data received from different
sources to detect dangerous events on the road. Their adopted cautious
operator is an extension of the Demper-Shafer theory that is known
to be superior in handling data come from dependent sources. 

A combined use of entity-centric and data-centric misbehavior detection
scheme makes use of both the trust level of vehicles and the consistency
of received data to detect misbehaving vehicles and filter out incorrect
messages. Works adopting the combined scheme are limited. In \cite{Dhurandher14},
Dhurandher proposed a security algorithm using both node reputation
and data plausibility checks to protect the network against attacks.
The reputation value is obtained by both direct monitoring and indirect
recommendation from neighbors; and the data consistency check is conducted
by comparing the received data with the sensed data by the vehicle's
own sensors. In \cite{Li16}, Li et al. proposed an attack-resistant
trust management scheme to evaluate the trustworthiness of both data
and vehicles in VANETs, and to detect and cope with malicious attacks.
They adopted the Dempster-Shafer theory to combine the data received
from different sources, and then used this combined result to update
the trust value of vehicles. 

In summary, all the above works on security issues in vehicular networks
focused on trust model establishment, trust model management, or methods
to assess data from different sources to check their consistency,
with a goal of detecting misbehaving nodes in the network. Our work
is different from theirs in that we focus on theoretically characterizing
the probability of correct message reception, and evaluate the impact
of network topology and distribution of malicious vehicles on the
probability. 

\section{System Model and Problem Formation\label{sec:System-Model-and}}

\subsection{Network Model}

We consider a vehicular ad hoc network on a highway with bi-directional
traffic flows. Vehicles in both directions are distributed randomly
following Poisson point processes \cite{Wis2007,Reis14} with spatial
densities $\rho_{1}$ and $\rho_{2}$ respectively. As a ready consequence
of the superposition property of Poisson processes \cite{Nelson95},
all vehicles on the highway are also Poissonly distributed with density
$\rho=\rho_{1}+\rho_{2}$. In actual road networks, there may be multiple
lanes in each direction. Considering that the width of a lane is typically
small compared with the transmission range of vehicles, we ignore
the road width and model multiple lanes in the same direction as one
lane \cite{Wis2007,Abboud14}. 

\subsection{Wireless Communication Model}

We consider a general wireless connection model \cite{Zhang14}, where
a receiver separated by a Euclidean distance $x$ from a transmitter
receives the message successfully with a probability $g(x)$, independent
of transmissions by other transmitter-receiver pairs. There are two
constraints on $g(x)$: 1) it is a monotonic non-increasing function
of $x$ and 2) $\lim_{x\rightarrow\infty}g(x)=0$. This general wireless
connection model includes a number of widely-used wireless connection
models as its special cases. For instance, when $g(x)$ assumes the
following form

\begin{equation}
g(x)=\begin{cases}
1, & 0<x\leq r\\
0, & x>r
\end{cases},
\end{equation}
it becomes the widely known unit disk model where a pair of wireless
nodes are directly connected when their Euclidean distance is smaller
than or equal to a threshold $r$, known as the transmission range.
Alternatively, when $g(x)$ takes the following form, 

\begin{equation}
g(x)=\frac{1}{2}\left(1-\text{erf}\left(\frac{10\alpha\log_{10}\left(\frac{x}{r}\right)}{\sqrt{2\sigma^{2}}}\right)\right),
\end{equation}
it becomes another widely known log-normal connection model \cite{Mao2017,Mao2012,Mao2013},
where $\alpha$ is the path loss exponent, $\sigma$ is the standard
deviation and $r$ is the equivalent transmission range when $\sigma=0$. 

We consider a network with a sufficiently large vehicular density
such that the generated vehicular network is a connected network \cite{Mao2017}.
Besides, broadcast transmission is adopted so that each message can
be received by multiple vehicles to increase the number of redundant
data forwarding paths and reduce the message dissemination time. Furthermore,
we assume that time is divided into time slots of equal length $\tau,$
and $\tau$ is sufficiently small such that we can regard vehicles
as almost stationary during each time slot. After the message dissemination
process begins, at each time slot, a vehicle among the set of vehicles
that 1) have received at least one message and 2) are yet to transmit
the message, is randomly chosen to broadcast its received message.
Such broadcast protocol can be readily implemented in a distributed
manner by having each vehicle waits a random amount of time identically
and independently distributed following an exponential distribution
before transmitting its received message. Each vehicle only transmits
its received message once. Note that the radio propagation speed is
much faster than the moving speed of vehicles \cite{Zijie14}. Therefore,
we ignore the information propagation delay in this paper and assume
that during the message dissemination process, the topology of the
vehicular network remains unchanged. 

\subsection{Malicious Vehicle distribution and Data Fusion Rule}

We assume that vehicles along the highway can be classified into two
categories: \textit{normal vehicles,} which behave normally and will
forward the received message without any alteration, and \textit{malicious
vehicles, }\textit{\emph{which}}\emph{ }may tamper the received message
and alter its content. Beside, we assume that the probability of each
vehicle being a malicious vehicle is $p_{m}$, independent of the
event that another distinct vehicle is a malicious vehicle. We further
assume that the malicious vehicles act in a distributed manner and
there is no central coordination among malicious vehicles. As a consequence
of the assumption, each malicious vehicle simply modifies the received
message without evaluation of the true content of the message. 

Following the broadcast dissemination scheme considered in the paper,
each vehicle is likely to receive multiple copies of a message from
different vehicles before it broadcasts the message. Due to the existence
of malicious vehicles, the received messages may not be the same.
For example, one vehicle may detect a traffic incident and generate
a message alerting other vehicles but this message may be modified
by a malicious vehicle. In the situation of conflicting messages being
received, a\textit{ majority voting} rule is employed by each vehicle
to fuse their received messages. That is, the normal vehicle will
broadcast the message in agreement with the most number of vehicles
and discard the message conflicting with majority opinion, and the
malicious vehicle will broadcast the message conflicting with the
majority opinion. When a tie occurs, all the vehicles will randomly
choose one of the two messages (true or false message) with equal
probability to broadcast. The simplicity of the majority voting rule
allows us to\textcolor{black}{{} focus on the topological impact of
vehicular networks on the correct message delivery. It is part of
our future work plan to investigate the optimum fusion rule for highly
dynamic vehicular networks.} 

\subsection{Problem Formation\label{subsec:Problem-Formation}}

Given the aforementioned background, we are now ready to give a formal
definition of the problem considered in this paper. 

\begin{figure}[t]
\centering{}\includegraphics[width=8.5cm]{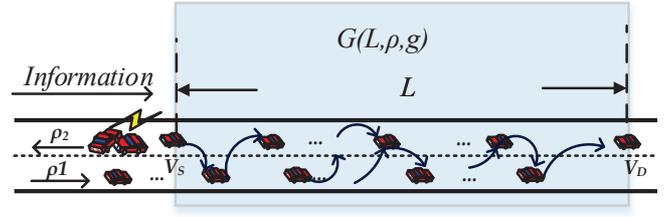} \caption{An illustration of the sub-network we focused on in this work, which
starts from the location of vehicle $V_{S}$ and ends at the location
of destination vehicle $V_{D}$.\label{Fig: System model} }
\end{figure}

Consider a vehicle, termed the source vehicle $V_{S}$, detects an
accident in front of it and wants to deliver a warning message to
vehicles traveling in the same direction as $V_{S}$ and behind $V_{S}$
in that direction. Designate the location of $V_{S}$ at the time
instant when it broadcasts the message as the origin, and the direction
of information propagation (in the opposite direction of the travel
direction of $V_{S}$) as the positive direction. We want to investigate
the probability that a vehicle, termed the destination vehicle $V_{D}$,
located at distance $L$ away from $V_{S}$ can receive the message
of $V_{S}$ correctly. We denote by $G(L,\rho,g)$ the sub-network
we focus on, which is within the road segment $(0,L)$, and with vehicular
density $\rho$ and a wireless connection model $g$. See Fig. \ref{Fig: System model}
for an illustration. 

Two kinds of messages are considered in this paper, $+1$ represents
the true message (e.g., road is congested) and $-1$ represents the
false message (e.g., road is not congested). Here we assume that the
source vehicle $V_{S}$ is a normal vehicle, namely, the message broadcast
by the source vehicle $V_{S}$ is true. For malicious vehicles, as
there are no central coordination among them, there is no way for
a malicious vehicle to know the true content of the message. Therefore,
it is assumed that a malicious vehicle simply modify the content of
whatever message it receives (against the outcome of the majority
voting rule), i.e., changing $+1$ to $-1$ and $-1$ to $+1$.

Finally, the destination vehicle $V_{D}$ conducts its majority voting
process after it has received all messages, or equivalently after
no further message is received after a long time period. Denote by
$M_{D}$ the concluded message after $V_{D}$ has completed its data
fusion. In this paper, we are interested in investigating the probability
that the destination vehicle $V_{D}$ receives the correct message,
denoted by $P_{succ}$, which can be expressed as follows:
\begin{equation}
P_{succ}=\Pr(M_{D}=1)\label{eq:definition of Psuc}
\end{equation}

\section{Theoretical Analysis \label{sec:Theoretical-Analysis }}

In this section, we will present our analysis on the probability that
the destination vehicle receives the message correctly. 

From the definition of the probability of correct message reception,
which is given in \eqref{eq:definition of Psuc}, $P_{succ}$ can
be expressed as follows as an easy consequence of the total probability
theorem:
\begin{align}
P_{succ}= & \Pr(M_{D}=1)\nonumber \\
= & \sum_{n=1}^{\infty}\Pr(M_{D}=1|N=n)\Pr(N=n)\label{eq:cal_Psuc}
\end{align}
where $N$ denotes the random number of vehicles located in the sub-network
$G(L,\rho,g)$. Due to the Poisson distribution of vehicles, we have
\begin{equation}
\Pr(N=n)=\frac{(\rho L)^{n}e^{-\rho L}}{n!}.\label{eq:Poisson distribution}
\end{equation}

Recall that in our system, the source vehicle $V_{S}$ located at
the origin broadcasts its message first. After that, at each time
slot, a vehicle among the set of vehicles having received at least
one message \textit{and} having not broadcast its message is randomly
chosen to broadcast. Denote by $V_{i}$ the $i$th vehicle that broadcasts
the message and denote its location by $Y_{i}$, where $Y_{i}\in\left(0,L\right),i=1,2,...n$
is a random variable representing the location of the $i$th vehicle
broadcasting its message. We designate the source vehicle $V_{S}$
as the $0$th broadcast vehicle and its location is $y_{0}=0$. It
follows that the destination vehicle $V_{D}$ then becomes the $(n+1)$th
broadcast vehicle. Using the total probability theorem, the conditional
probability that the destination vehicle $V_{D}$ receives the correct
message (after its fusion), given there are $N=n$ vehicles located
in the sub-network $G(L,\rho,g)$, can be calculated by
\begin{align}
 & \Pr(M_{D}=1|N=n)\nonumber \\
= & \int_{0}^{L}\cdots\int_{0}^{L}\int_{0}^{L}\Pr(M_{D}=1|Y_{1}=y_{1},Y_{2}=y_{2},...Y_{n}=y_{n})\nonumber \\
 & \;\;\times f_{Y_{1},Y_{2},...Y_{n}}(y_{1},y_{2},...y_{n})dy_{1}dy_{2}...dy_{n}\label{eq:Psuc condition on N}
\end{align}
where $f_{Y_{1},Y_{2},...Y_{n}}(y_{1},y_{2},...y_{n})$ is the joint
distribution (probability density function) of the locations of the
$1$st, $2$nd, $...$, and $n$th broadcast vehicles.

Combining \eqref{eq:cal_Psuc} - \eqref{eq:Psuc condition on N},
it can be shown that to obtain the correct message reception probability
$P_{succ}$, it remains to calculate the conditional probability that
the destination vehicle $V_{D}$ receives the message correctly given
that the $i$th broadcast vehicle is located at $y_{i},i=1,2,...n$,
i.e., $\Pr(M_{D}=1|Y_{1}=y_{1},Y_{2}=y_{2},...Y_{n}=y_{n})$, and
the joint distribution of the locations of the $1$st, $2$nd, $...$,
and $n$th broadcast vehicles, i.e., $f_{Y_{1},Y_{2},...Y_{n}}(y_{1},y_{2},...y_{n})$.
In the following, we will calculate these two terms separately. 

\subsection{Calculation of $Pr(M_{D}=1|Y_{1}=y_{1},Y_{2}=y_{2},...Y_{n}=y_{n})$}

Denote by $h(y_{i}),i=0,1,...n$ the indicator function that represents
whether the destination vehicle $V_{D}$ receives the message sent
by the $i$th broadcast vehicle $V_{i}$ located at $Y_{i}=y_{i}$.
Following the general wireless connection model considered in the
paper, it can be readily shown that
\begin{equation}
h(y_{i})=\begin{cases}
1, & g\left(L-y_{i}\right)\\
0, & 1-g\left(L-y_{i}\right)
\end{cases},i=0,1,...n.\label{eq:h(y_i)}
\end{equation}

Denote by $M_{i}$ the message broadcast by the $i$th broadcast vehicle
$V_{i}$ located at $y_{i}$, $i=0,1,...n$. It follows that $M_{0}=1$
as we regard the source vehicle $V_{S}$ is a normal vehicle that
broadcasts the true message, and each $M_{i},i=1,...n$ is a binary
random variable taking value from $\left\{ +1,-1\right\} $. Assuming
the majority voting rule, the conditional probability that the destination
vehicle $V_{D}$ receives the message correctly given that the $i$th
broadcast vehicle is located at $y_{i}$, $i=1,...n$, can be calculated
by:
\begin{align}
 & \Pr(M_{D}=1|Y_{1}=y_{1},Y_{2}=y_{2},...Y_{n}=y_{n})\nonumber \\
= & \Pr\left(\sum_{i=0}^{n}M_{i}h(y_{i})>0\right)+\frac{1}{2}\Pr\left(\sum_{i=0}^{n}M_{i}h(y_{i})=0\right)\nonumber \\
= & \sum_{j=1}^{2^{n+1}}\left[\Pr\left(\sum_{i=0}^{n}M_{i}h^{j}(y_{i})>0\right)\Pr\left(\mathbf{h=h}^{j}\right)\right]\nonumber \\
 & +\frac{1}{2}\sum_{j=1}^{2^{n+1}}\left[\Pr\left(\sum_{i=0}^{n}M_{i}h^{j}(y_{i})=0\right)\Pr\left(\mathbf{h=h}^{j}\right)\right]\nonumber \\
= & \sum_{j=1}^{2^{n+1}}\Biggl\{\Pr\left(\sum_{i=0}^{n}M_{i}h^{j}(y_{i})>0\right)\times\nonumber \\
 & \left[\prod_{i=0}^{n}\left[g\left(L-y_{i}\right)h^{j}(y_{i})+\left(1-g\left(L-y_{i}\right)\right)\left(1-h^{j}(y_{i})\right)\right]\right]\Biggr\}\nonumber \\
 & +\frac{1}{2}\sum_{j=1}^{2^{n+1}}\Biggl\{\Pr\left(\sum_{i=0}^{n}M_{i}h^{j}(y_{i})=0\right)\times\nonumber \\
 & \left[\prod_{i=0}^{n}\left[g\left(L-y_{i}\right)h^{j}(y_{i})+\left(1-g\left(L-y_{i}\right)\right)\left(1-h^{j}(y_{i})\right)\right]\right]\Biggr\}\label{eq:cal of Psuc given each yi}
\end{align}
where the vector $\mathbf{h}$ is defined by $\mathbf{h}=\left\{ h(y_{0}),h(y_{1}),...h(y_{n}):h(y_{i})\in\{1,0\},1\leq i\leq n\right\} $,
and the first step follows from the rule of majority voting, particularly
noting that when a tie occurs, the destination vehicle will make a
decision randomly with equal probability. The second step is obtained
by using the total probability theorem on $\mathbf{h}$. Note from
\eqref{eq:h(y_i)} that each $h(y_{i}),i=0,1,...n$ is a binary random
variable. Therefore, the vector $\mathbf{h=}\left\{ h(y_{0}),h(y_{1}),...h(y_{n})\right\} $
can have $2^{n+1}$ possible values and we let $\mathbf{h=h}^{j},\;j=1,2,...2^{n+1}$
represents each possible value. The third step follows by plugging
$\Pr\left(\mathbf{h=h}^{j}\right)=\prod_{i=0}^{n}\left[g\left(L-y_{i}\right)h^{j}(y_{i})+\left(1-g\left(L-y_{i}\right)\right)\left(1-h^{j}(y_{i})\right)\right]$,
which readily results from the definition of each $h(y_{i}),i=0,1,...n$
given as \eqref{eq:h(y_i)}. 

From \eqref{eq:cal of Psuc given each yi}, to calculate $\Pr(M_{D}=1|Y_{1}=y_{1},Y_{2}=y_{2},...Y_{n}=y_{n})$,
it remains to calculate the two terms $\Pr\left(\sum_{i=0}^{n}M_{i}h^{j}(y_{i})>0\right)$
and $\Pr\left(\sum_{i=0}^{n}M_{i}h^{j}(y_{i})=0\right)$ given each
fixed $\mathbf{h}^{j}=\left\{ h^{j}(y_{0}),h^{j}(y_{1}),...h^{j}(y_{n})\right\} $,
$j=1,2,...2^{n+1}$. Using the joint distribution of $M_{1},$ $M_{2}$,
$...$ $M_{n}$, $\Pr\left(M_{1}=m_{1},M_{2}=m_{2},...M_{n}=m_{n}\right)$,
the above two terms can be obtained as follows: 
\begin{align}
 & \Pr\left(\sum_{i=0}^{n}M_{i}h^{j}(y_{i})>0\right)\nonumber \\
= & \sum_{\sum_{i=0}^{n}m_{i}h^{j}(y_{i})>0}\Pr\left(M_{0}=m_{0},M_{1}=m_{1},...M_{n}=m_{n}\right)\nonumber \\
= & \sum_{h^{j}(0)+\sum_{i=1}^{n}m_{i}h^{j}(y_{i})>0}\Pr\left(M_{1}=m_{1},...M_{n}=m_{n}\right),\label{eq:Pr(sum Mi>0)}
\end{align}
and 
\begin{align}
 & \Pr\left(\sum_{i=0}^{n}M_{i}h^{j}(y_{i})=0\right)\nonumber \\
= & \sum_{\sum_{i=0}^{n}m_{i}h^{j}(y_{i})=0}\Pr\left(M_{0}=m_{0},M_{1}=m_{1},...M_{n}=m_{n}\right)\nonumber \\
= & \sum_{h^{j}(0)+\sum_{i=1}^{n}m_{i}h^{j}(y_{i})=0}\Pr\left(M_{1}=m_{1},...M_{n}=m_{n}\right).\label{eq:Pr(sum Mi=00003D0)}
\end{align}

According to the chain rule of probability, it can be readily obtained
that the joint distribution of $M_{1},$$M_{2}$ $...$ $M_{n}$,
$\Pr\left(M_{1}=m_{1},M_{2}=m_{2},...M_{n}=m_{n}\right)$ is given
by
\begin{align}
 & \Pr\left(M_{1}=m_{1},M_{2}=m_{2},...M_{n}=m_{n}\right)\nonumber \\
= & \Pr\left(M_{n}=m_{n}|M_{n-1}=m_{n-1},...M_{2}=m_{2},M_{1}=m_{1}\right)\times\nonumber \\
 & \Pr\left(M_{n-1}=m_{n-1}|M_{n-2}=m_{n-2},...M_{2}=m_{2},M_{1}=m_{1}\right)\nonumber \\
 & \times...\times\Pr\left(M_{2}=m_{2}|M_{1}=m_{1}\right)\Pr(M_{1}=m_{1}).\label{eq:Joint distribution of Mi}
\end{align}

Note that the message fusion result of vehicle $V_{i}$ is dependent
on the messages $M_{0},$ $M_{1},$ $...$ $M_{i-1}$ broadcast by
vehicles $V_{S},V_{1},...V_{i-1}$. Therefore, the conditional distribution
of each $M_{i}$, $i=1,2,...n$ given $M_{1}=m_{1},...M_{i-1}=m_{i-1}$
can be obtained as follows:
\begin{align}
 & \Pr\left(M_{i}=1|M_{0}=1,M_{1}=m_{1},...M_{i-1}=m_{i-1}\right)\nonumber \\
= & \Pr\left(1+\sum_{j=1}^{i-1}\left(m_{j}\cdot g(y_{i}-y_{j})\right)>0\right)(1-p_{m})\nonumber \\
 & +\Pr\left(1+\sum_{j=1}^{i-1}\left(m_{j}\cdot g(y_{i}-y_{j})\right)<0\right)p_{m}\nonumber \\
 & +\frac{1}{2}\Pr\left(1+\sum_{j=1}^{i-1}\left(m_{j}\cdot g(y_{i}-y_{j})\right)=0\right)\label{eq:conditional distribution of Mi}
\end{align}
and 
\begin{align}
 & \Pr(M_{i}=-1|M_{0}=1,M_{1}=m_{1},...M_{i-1}=m_{i-1})\nonumber \\
= & 1-\Pr\left(M_{i}=1|M_{0}=1,M_{1}=m_{1},...M_{i-1}=m_{i-1}\right),\label{eq:Pr(M_i=00003D-1)}
\end{align}
where the three terms in \eqref{eq:conditional distribution of Mi}
are the probabilities that vehicle $V_{i}$ broadcasts message $+1$
under three different cases: $1+\sum_{j=1}^{i-1}\left(m_{j}\cdot g(y_{i}-y_{j})\right)>0$,
$1+\sum_{j=1}^{i-1}\left(m_{j}\cdot g(y_{i}-y_{j})\right)<0$, and
$1+\sum_{j=1}^{i-1}\left(m_{j}\cdot g(y_{i}-y_{j})\right)=0$ separately.
Using the first case $1+\sum_{j=1}^{i-1}\left(m_{j}\cdot g(y_{i}-y_{j})\right)>0$
as an example to illustrate: when $1+\sum_{j=1}^{i-1}\left(m_{j}\cdot g(y_{i}-y_{j})\right)>0$,
vehicle $V_{i}$ would conclude from its majority voting process that
the majority opinion of the message is $+1$. Considering each vehicle
has probability $p_{m}$ to modify the message (being a malicious
vehicle), therefore, the probability for the vehicle to broadcast
the correct concluded message (from the majority voting process) $+1$
would be $1-p_{m}$, which leads to the term $\Pr\left(1+\sum_{j=1}^{i-1}\left(m_{j}\cdot g(y_{i}-y_{j})\right)>0\right)(1-p_{m})$.
Specifically, from \eqref{eq:conditional distribution of Mi}, when
$i=1$ we have 

\begin{equation}
\Pr\left(M_{1}=1\right)=1-p_{m},\label{eq:Probability of M1=00003D1}
\end{equation}
 and 
\begin{equation}
\Pr\left(M_{1}=-1\right)=p_{m},\label{eq:probability of M1=00003D-1}
\end{equation}
which can also be readily obtained as the $1$st broadcast vehicle
only receives the true message from the source vehicle.

Combining \eqref{eq:Joint distribution of Mi} - \eqref{eq:Pr(M_i=00003D-1)},
we can obtain the joint distribution of $M_{1},$ $M_{2},$ $...$
$M_{n}$, $\Pr\left(M_{1}=m_{1},M_{2}=m_{2},...M_{n}=m_{n}\right)$.
Plugging this joint distribution in \eqref{eq:Pr(sum Mi>0)} and \eqref{eq:Pr(sum Mi=00003D0)},
the two terms $\Pr\left(\sum_{i=0}^{n}M_{i}h^{j}(y_{i})>0\right)$
and $\Pr\left(\sum_{i=0}^{n}M_{i}h^{j}(y_{i})=0\right)$ in \eqref{eq:cal of Psuc given each yi}
can be obtained, which in turn leads to the result of $\Pr(M_{D}=1|Y_{1}=y_{1},Y_{2}=y_{2},...Y_{n}=y_{n})$. 

\subsection{Calculation of $f_{Y_{1},Y_{2},...Y_{n}}(y_{1},y_{2},...y_{n})$}

Let $K_{m},m=0,1,...n$ be the set of vehicles in the sub-network
$G(L,\rho,g)$ which have received at least one message after the
$m$th broadcast vehicle $V_{m}$ has broadcast its messages. Given
the location of the $i$th broadcast vehicle $V_{i}$ as $Y_{i}=y_{i},i=0,1,...m$,
a vehicle located at $x,\;x\neq y_{i},i=0,1,...m$ belongs to $K_{m}$
implies that it connects to at least one vehicle that are located
at $y_{0},y_{1},...y_{m}$, which has the probability $1-\prod_{i=0}^{m}\left(1-g\left(\vert x-y_{i}\vert\right)\right)$.
Note that the $(m+1)$th broadcast vehicle $V_{m+1}$ is randomly
chosen from the vehicle set $K_{m}\setminus\left\{ V_{1},...V_{m}\right\} $,
therefore, given each $Y_{i}=y_{i},i=1,2,...m$, the location of $(m+1)$th
broadcast vehicle $Y_{m+1}$ has the conditional probability density
function as follows: 
\begin{align}
 & f_{Y_{m+1}|Y_{1},Y_{2},...Y_{m}}(x|y_{1},y_{2},...y_{m})\nonumber \\
= & \frac{1-\prod_{i=0}^{m}\left(1-g\left(\vert x-y_{i}\vert\right)\right)}{\int_{0}^{L}\left[1-\prod_{i=0}^{m}\left(1-g\left(\vert x-y_{i}\vert\right)\right)\right]dx},\;\;m=0,1,...n-1\label{eq:conditional distribution of each yi}
\end{align}
Eq. \eqref{eq:conditional distribution of each yi} is valid when
$x\neq y_{i},i=1,2,...m$ as we assume each vehicle only broadcasts
once. Particularly, when $m=0$, we have the probability density function
of the $1$st broadcast vehicle's location
\begin{equation}
f_{Y_{1}}(x)=\frac{g(x)}{\int_{0}^{L}g(x)dx}.
\end{equation}

As an easy consequence of the chain rule of probability, the joint
distribution of $Y_{1},Y_{2},...Y_{n}$ can be obtained as follows:
\begin{align}
 & f_{Y_{1},Y_{2},...Y_{n}}(y_{1},y_{2},...y_{n})\nonumber \\
= & f_{Y_{n}|Y_{n-1},...Y_{2},Y_{n-1}}(y_{n}|y_{n-1},...,y_{2},y_{1})\nonumber \\
 & \times f_{Y_{n-1}|Y_{n-2},..Y_{2},Y_{1}}(y_{n-1}|y_{n-2},...,y_{2}y_{1})\nonumber \\
 & \times f_{Y_{n-2}|Y_{n-3},...Y_{2},Y_{1},}(y_{n-2}|y_{n-3},...y_{2},y_{1})\times...\nonumber \\
 & \times f_{Y_{2}|Y_{1}}(y_{2}|y_{1})\times f_{Y_{1}}(y_{1}),\label{eq:cal of joint distribution of yi}
\end{align}
where each conditional distribution in \eqref{eq:cal of joint distribution of yi}
is given by \eqref{eq:conditional distribution of each yi}.

\section{Simulation and Discussion\label{sec:Simulation-and-Discussion}}

In this section, numerical and simulation results are shown to discuss
the relationship between the probability of correct message reception
and its major performance-impacting parameters. Specifically, we adopt
the unit disk model and the log-normal connection model as two special
cases of the general wireless connection model respectively in the
simulation. For the unit disk model, we set the transmission range
$r=250$m (typical radio range using DSRC \cite{haibo14}), and for
the log-normal connection model, we set the the path loss exponent
$\alpha=2$ and the standard deviation $\sigma=4$ \cite{Zhang14}.
Each simulation is repeated 5000 times and the average value is shown
in the plot.

\begin{figure}
\centering{\includegraphics[width=7.5cm]{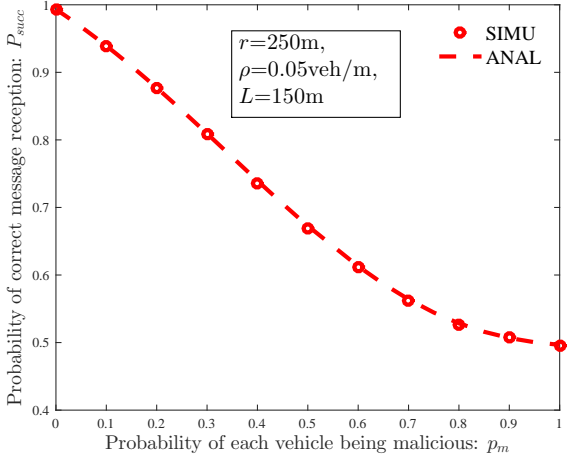}}

\caption{A comparison between analytical result and simulation result. }

\label{SIMU_ANAL}
\end{figure}

Fig. \ref{SIMU_ANAL} shows a comparison between the analytical result
and the simulation result assuming the unit disk model, and shows that the analytical result matches
very well with the simulation result. 

\begin{figure}
\centering{\includegraphics[width=7.5cm]{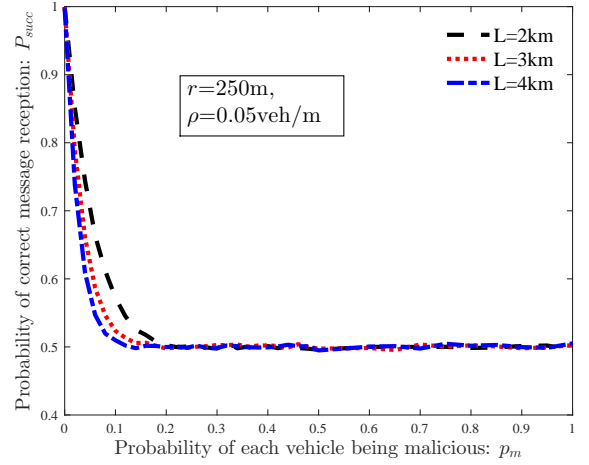}}

\caption{The relationship between the probability of correct message reception
$P_{succ}$ and the probability of each vehicle being malicious $p_{m}$
assuming the unit disk model, with different distance $L$ between
the source vehicle and the destination vehicle. }

\label{UDM_Psucc_Pm_L}
\end{figure}

\begin{figure}
\centering{\includegraphics[width=7.5cm]{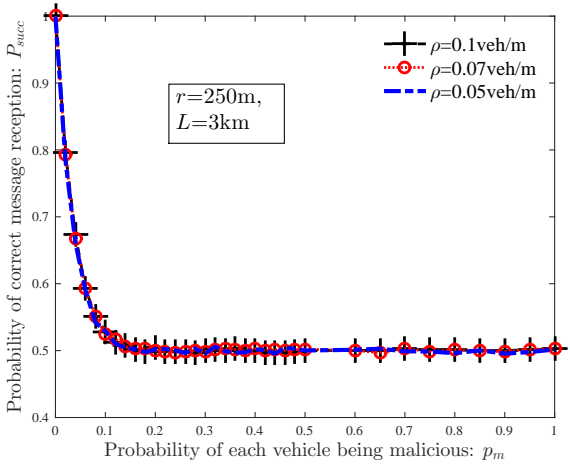}}

\caption{The relationship between the probability of correct message reception
$P_{succ}$ and the probability of each node being malicious $p_{m}$
assuming the unit disk model, with different vehicular density $\rho$.}

\label{UDM_Psucc_Pm_rho}
\end{figure}

Fig. \ref{UDM_Psucc_Pm_L} and Fig. \ref{UDM_Psucc_Pm_rho} show the
relationship between the probability of correct message reception
$P_{succ}$ and the probability of each vehicle being malicious $p_{m}$
assuming the unit disk model, under different distance $L$ between
the source vehicle and the destination vehicle, and under different
vehicular density $\rho$ respectively. Specifically, we can see that
$P_{succ}=1$ when $p_{m}=0$, which corresponds to the case that
all vehicles are normal vehicles; when $p_{m}$ is small, $P_{succ}$
decreases sharply with an increase of $p_{m}$ and decreases to its
minimum value ($0.5$ in our system) when $p_{m}$ is larger than
a certain threshold $p_{th}$, e.g., $p_{th}=0.2$ when $L=3$km and
$\rho=0.05$veh/m. Beyond that threshold, a further increase in $p_{m}$
has little impact on $P_{succ}$. This can be explained by the fact
that when $p_{m}<p_{th}$, the number of malicious vehicles in the
network is small. Therefore, an increase in $p_{m}$ will largely
increase the number of malicious vehicles, which consequently, leads
to a sharp decrease in the probability of correct message reception.
When $p_{m}$ is larger than its threshold, malicious vehicles play
dominant roles in the majority voting scheme. In this case, for any
vehicle in the network, the outcome of its message fusion result will
be incorrect. The minimum value of $P_{succ}=0.5$ is due to the fact
that malicious vehicles in our network simply modify the received
message without evaluation of the true content of the message. Therefore,
when $p_{m}$ is larger than its threshold, the message transmitted
in the network will move between $+1$ and $-1$ back and forth, leading
to the occurrence that $P_{succ}$ converges to 0.5 instead of 0. 

Fig. \ref{UDM_Psucc_Pm_L} shows that given a fixed vehicular density,
when $p_{m}<p_{th}$, a larger distance $L$ between the source vehicle
and the destination vehicle will lead to a smaller $P_{succ}$. This
is due to the fact that other things being equal, a larger $L$ implies
a larger number of malicious vehicles participating in tampering the
message transmitted from the source vehicle to the destination vehicle.
As a consequence, it leads to a smaller $P_{succ}$.

Fig. \ref{UDM_Psucc_Pm_rho} shows that in our system, a larger vehicular
density $\rho$ has little impact on $P_{succ}$. Intuitively, a larger
$\rho$ will lead to a larger $P_{succ}$ due to the fact that a larger
$\rho$ implies a larger number of messages received by each vehicle,
which is beneficial for vehicles to conduct data consistency checks.
Therefore, when the traffic density increases, the message fusion
result of each vehicle will be more accurate. Consequently, other
things being equal, the probability of correct message reception $P_{succ}$
will increase. However, when a vehicle is randomly chosen among the
set of vehicles that have received at least one message to broadcast,
it may not have received a sufficient number messages from other vehicles
to conduct a robust data fusion. This follows that even with an increase
in traffic density $\rho$, the message fusion result of each broadcast
vehicle does not improve. Therefore, a larger vehicular density $\rho$
has little impact on the $P_{succ}$. 

\begin{figure}
\centering{\includegraphics[width=7.5cm]{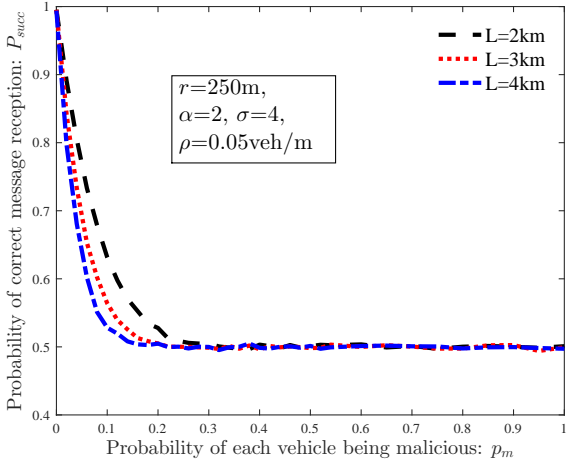}}

\caption{The relationship between the probability of correct message reception
$P_{succ}$ and $p_{m}$ assuming the log-normal connection model,
with different distance $L$ between the source vehicle and the destination
vehicle. }

\label{LSM_Psucc_Pm_L}
\end{figure}

\begin{figure}
\centering{\includegraphics[width=7.5cm]{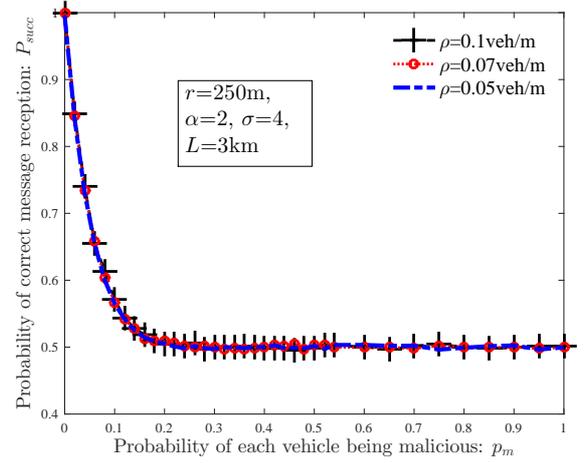}}

\caption{The relationship between the probability of correct message reception
$P_{succ}$ and $p_{m}$ assuming the log-normal connection model,
with different vehicular density $\rho$.}

\label{LSM_Psucc_Pm_rho}
\end{figure}

Fig. \ref{LSM_Psucc_Pm_L} and Fig. \ref{LSM_Psucc_Pm_rho} show the
relationship between the probability of correct message reception
$P_{succ}$ and the probability of each vehicle being malicious $p_{m}$
assuming the log-normal connection model, under different distance
$L$ between the source vehicle and the destination vehicle, and under
different vehicular density $\rho$ respectively. We can see that
with the increase of $p_{m}$ from $0$ to $1$, the trend of $P_{succ}$
is the same as that assuming the unit disk model. Therefore, we omit
the duplicate discussion here.

\begin{figure}
\centering{\includegraphics[width=7.5cm]{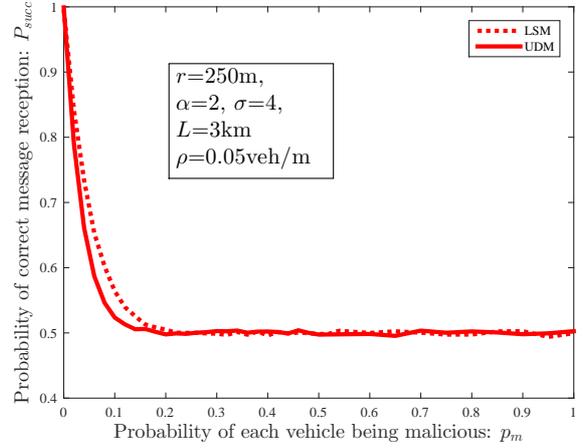}}

\caption{A comparison between the probability of correct message reception
$P_{succ}$ achieved assuming the unit disk model and that assuming
the log-normal connection model. }

\label{LSM_UDM}
\end{figure}

\textcolor{black}{Fig. \ref{LSM_UDM} gives a comparison of the} correct
message reception probability $P_{succ}$ achieved\textcolor{black}{{}
assuming the unit disk model (labeled as UDM) and that achieved assuming
the log-normal connection model (labeled as LSM). It is shown that}
when $p_{m}<p_{th}$,\textcolor{black}{{} the system assuming the log-normal
connection model has a slightly higher }correct message reception
probability $P_{succ}$\textcolor{black}{{} than that assuming the unit
disk model. The reason behind this phenomenon is that the log-normal
connection model introduces a Gaussian variation of the transmission
range around the mean value, which implies a higher chance for the
vehicles to be connected to other vehicles separated further away.
Therefore, other things being equal, each broadcast vehicle assuming
the log-normal connection model can receive more copies of a message
from other vehicles than that assuming the unit disk model, which
leads to a better message fusion result for each vehicle and consequently,
results in a higher }correct message reception probability. 

\section{Conclusions \label{sec:Conclusion-and-Future}}

This paper studied a vehicular ad hoc network where a certain fraction
of vehicles are malicious vehicles and these malicious vehicles are
distributed randomly in the network. Furthermore, there is no central
coordination among these malicious vehicles and consequently a malicious
vehicle simply modify its received message irrespective of its true
value. An analytical framework is developed to model the process of
secure message dissemination in the network, and the probability that
a vehicle, located at a fixed distance away from the source vehicle,
can receive the message correctly is obtained. Simulations were conducted
to establish the accuracy of the analytical results and demonstrate
that the probability of correct message delivery reduces to its minimum
after the proportion of malicious vehicles in the network increases
beyond a threshold. Besides, a smaller distance between the destination
vehicle and the source vehicle will lead to a larger probability of
correct message reception. Our results may provide insight on the
design of security mechanisms, particularly secure routing algorithms
and topology control algorithms, to enhance secure message dissemination
in highly dynamic vehicular networks.

\end{document}